\documentclass{ws-rv9x6}
\usepackage{subfigure}   % required only when side-by-side / subfigures are used
\usepackage{ws-rv-thm}   % comment this line when `amsthm / theorem / ntheorem` package is used
\usepackage{ws-rv-van}   % numbered citation & references (default)
\usepackage{caption}

\makeindex
\begin{document}

\chapter[The Essential Role of Thermodynamics in metabolic network modeling]{The Essential Role of Thermodynamics in metabolic network modeling: physical insights and computational challenges}\label{ra_ch1}

\author[A. De Martino, D. De Martino and E. Marinari]{A. De Martino$^{1,4}$ , D. De Martino$^2$ and E. Marinari$^3$}
%\index[aindx]{Author, F.} % or \aindx{Author, F.}
%\index[aindx]{Author, S.} % or \aindx{Author, S.}

\address{$^1$ Soft \& Living Matter Lab, CNR-NANOTEC, Rome (Italy) \\
$^2$ Institute of Science and Technology Austria,
Am Campus 1, A-3400 Klosterneuburg (Austria) \\
and Jozef Stefan Institute, Jamova 39, SI-1000,  Ljubljana, Slovenia\\
$^3$ Sapienza  Universit\`a  di  Roma,  INFN  Sezione  di  Roma  1  and  CNR-NANOTEC,
UOS  di  Roma,  P.le  A.  Moro  2,  00185  Roma  (Italy) \\
$^4$ Italian Institute for Genomic Medicine, Turin (Italy)}

\begin{abstract}
Quantitative studies of cell metabolism are often based on large chemical reaction network models. A steady state approach is suited to analyze phenomena on the timescale of cell growth and  circumvents the problem of incomplete experimental knowledge on kinetic laws and parameters, but it shall be supported by a correct implementation of thermodynamic constraints. In this article we review the latter aspect  highlighting its computational challenges and  physical insights.  The simple introduction of Gibbs inequalities avoids the presence of unfeasible loops allowing for correct timescale analysis but leads to possibly non-convex feasible flux spaces, whose exploration needs efficient algorithms. We shorty review on the implementation of thermodynamics through variational principles in constraints based models of metabolic networks.           
\end{abstract}
%\markright{Customized Running Head for Odd Page} % default is Chapter Title.
\body

%\tableofcontents

%%%%%%%%%%%%%%%%%%%%%%%%%%%%%%%%%%%%%%%%%%%%%%%%%%%% adm
\section*{Introduction}

Because of its uniquely universal nature, thermodynamics has been linked to physiology from its very inception, both to  rationalize observations and to elucidate fundamental limits to physiological functions. With the advent of genetics and molecular biology, the discovery of the molecular mechanisms underlying physiology became the primary challenge. However, as the molecular actors and their interactions were mapped out at increasingly fine resolution, the focus gradually shifted on understanding their system-level organization \cite{kacser1973control}. And, perhaps unsurprisingly, it has become more and more clear that thermodynamic aspects are crucial for the emergent large-scale behaviour of these systems. Currently, renewed interest has flourished around the thermodynamics of cellular processes, only this time with the possibility of relying on a host of data at various scales for quantitative analyses \cite{Albertybook}.

In no area of physiology is thermodynamic analysis more central than in metabolic network modeling \cite{ataman2015heading}. In brief, metabolic networks encode for the set of chemical reactions that, in any cell, break down nutrients and harvest free energy to synthetize the macro-molecular building blocks essential to life (amino acids, nucleotides, fatty acids, etc.) and, ultimately, biomass. Their structures can be inferred by combining gene-enzyme-reactions associations with regulatory information and transcriptional data. The availability of detailed metabolic network reconstructions for a large number of organisms and cell types is, in our view, among the most significant successes that computational methods have reaped in biology to date \cite{patil2004use}. 

Building reliable and predictive dynamical models of metabolism based on this information is however challenging, mainly due to our vastly incomplete knowledge about intracellular enzyme kinetics, transport mechanisms and rate constants. On the other hand, non-equilibrium steady state approaches appear to be more feasible. Such methods are perhaps best represented by the broad class of computational schemes known as `constraint-based models' \cite{palssonbook, chembiop}. From a physical viewpoint, such  models should essentially rely on two ``Kirchhoff-type'' assumptions regarding (a) mass balance for chemical species (i.e. metabolic homeostasis) and (b) energy balance for reactions (i.e. thermodynamic feasibility of material fluxes) \cite{Beard:2002vn}. In the most basic setup, energy balance simply requires reaction fluxes at steady state to proceed downhill in free energy, in accordance with the second law of thermodynamics. Unfortunately, implementing this constraint in genome-scale models is drastically harder than enforcing the  stationarity of metabolite concentrations. Inclusion of thermodynamic constraints is however essential not only to obtain physically viable flux patterns, but also to highlight timescales and turnover rates and to allow for the estimation of metabolite concentrations. Needless to say, the range of applications of such results, from biotechnology to pharmacology, would be enormous.

Our main goal here is to present this problem and its multiple ramifications, which span from basic biochemistry to some fundamental algorithmic challenges, under a statistical physics lens. We shall discuss what, in practice, makes it so hard to solve efficiently, and review some of the alternative approaches that have been attempted. Finally, we will point to some recent developments that may hold some promising keys to finally unlock the puzzle of metabolic network thermodynamics at genome resolution.

%%%%%%%%%%%%%%%%%%%%%%%%%%%%%%%%%%%%%%%%%%%%%% ddm
\section*{Background} 
In the most simple setting  metabolism can be modeled   in terms of the dynamics of the chemical compounds concentration levels \cite{heinrichregulation}. Upon assuming well-mixing and neglecting noise, we still have a  large possibly non-linear dynamical system whose parameters   could be not  known in their entirety.
For a chemical reaction  network  in which $M$ metabolites participate in $N$ reactions  with the  stoichiometry encoded in a matrix $\mathbf{S}=\{S_{\mu i}\}$,  the concentrations $c_\mu$ change in time according to mass-balance equations
\begin{equation}
\dot{\mathbf{c}} = \mathbf{S \cdot v}
\end{equation}
where a component of the vector ${\bf v}$,  $v_i$, is the flux of the reaction $i$ that is in turn a (possibly unknown) function of the concentration levels  $v_i(\mathbf{c})$ (and several other parameters, like enzyme copy number, etc).
On the other hand, in order to analyze phenomena with timescales longer than diffusion and typical turnover times (like cell growth) it is possible to assume a steady state, i.e. a flux configuration satisfying 
\begin{equation}
\mathbf{S \cdot v}=0
\end{equation}
In so called constraints-based modeling, apart from mass balance, fluxes are bounded in certain ranges $v_r \in [v_{r}^{{\rm min}},v_{r}^{{\rm max}}]$ that take into account thermodynamic irreversibility, kinetic limits and physiological constraints.
The set of constraints
\begin{eqnarray}\label{eq3}
\mathbf{S \cdot v}=0, \nonumber \\
v_r \in [v_{r}^{{\rm min}},v_{r}^{{\rm max}}]
\end{eqnarray}  
defines  a convex closed set in the space of reaction fluxes: the polytope of feasible steady states. The productive capabilities of the network can be 
investigated computationally by maximizing suited linear objective functions in the aforementioned space \cite{majewski1990simple}, in particular the biomass growth itself (flux balance analysis \cite{Orth:2010if}, based on linear programming). On the other hand more generic inference problems can be afforded quite efficiently with Monte Carlo methods given the convexity of the space \cite{uniformell}.
It shall be noted that thermodynamics is implemented in a very simple way, i.e. by setting reaction reversibility, i.e. $f_{i}\geq 0$ for some reactions.
When flux bounds are not provided, it is customary to set them to an arbitrary  large number, $f_i \in \left[-C,C\right]$ (typically $C=10^3, 10^4$),  that, for a meaningful model, shall not influence the results.
On the other hand a more rigorous yet simple approach consists in postulating that  fluxes shall follow  a free energy gradient. If $f_i \neq 0$, then $f_i \Delta G_i < 0$ (Gibbs inequality) where $\Delta G_i$ is the free energy change of reaction $i$. The $\Delta G$'s can be written as the difference between the chemical potentials $g_\mu$ of products and substrates through the stoichiometric matrix $\Delta G_i = \sum_\mu S_{i \mu} g_\mu$. 
In terms of chemical potentials we have thus a system of linear inequalities ($\xi_{i \mu} = - sign(f_i) S_{i \mu}$) 
\begin{equation}
\sum_\mu \xi_{i \mu} g_\mu >0  \quad \forall i 
\end{equation}  
whose feasibility (existence of a solution $g_\mu$) is necessary for the thermodynamical feasibility of the flux configuration.
Even in this basic approach, the addition of free energy variables ($g_\mu$ and $\Delta G_i$) makes the problem  non-linear, in particular quadratic and possibly non-convex. On the other hand, upon conditioning on flux variables we can get useful hints from duality theorems of the alternative that characterize thermodynamic feasibility in terms of the unfeasibility of particular flux configurations \cite{de2013thermodynamics}. Specifically, according to the Gordan theorem we have that  system (4) has a feasible solution if and only if
\begin{equation}
\sum_i S_{i \mu} k_i  =0 \quad  k_i \geq 0  \quad \forall \mu     
\end{equation}
has no non-trivial solutions (unfeasible loops). Such a duality can be exploited in order to define efficient algorithms as we discuss in the following section.

%%%%%%%%%%%%%%%%%%%%%%%%%%%%%%%%%%%%%%%%%%% em
\section*{Relaxational Algorithms}  

Reconstruction of complete metabolic networks is becoming, thanks to new, modern and accurate experiments, a possible option for many simple organisms. What is important in the present context is that this reconstruction needs to be compatible with thermodynamic principles, and that implementing thermodynamic requirements can help in an accurate reconstruction of the network. This is sometimes a difficult computational problem, and we describe here an useful algorithm to implement thermodynamic consistency and use this to help in a (correct) network reconstruction \cite{de2012scalable}.

We will in this way gather information about Gibbs free energy and about their landscape, that will have to be compatible with the selected vector of reaction directions. We use stoichiometric information via a constructive algorithm  inspired by perceptron \cite{minover0} approach learning. In the method we use the a preliminary reconstruction of the  network structure to iteratively build up correlations between the chemical potentials of the chemical species,  until we reach a thermodynamically consistent profile.  The  algorithm is nicely scalable, and it can allow the crucial result of guaranteeing the feasibility of flux configurations, or of identifying and removing unfeasible cycles. The algorithm can also be useful to get an estimate of reaction affinities, and it can be used to derive bounds for concentrations.

We consider the Gibbs energy at temperature $T$ and volume $V$
\begin{equation}
    G \equiv E - P\; V-T\;S\;,
\end{equation}
where $E$ is the internal energy of the systems, $P$ its pressure and $S$ the entropy. Let us call $\delta_i$ the direction of chemical reaction $i$: $\delta_i=\pm 1$, i.e. the reaction can proceed in the "forward" direction or in the reverse direction. If $\Delta G_i$ is the Gibbs energy difference induced by reaction $i$ one needs that $\delta_i\,\Delta G_i \le 0\;\forall i$ (the Gibbs energy cannot increase in the direction where the chemical reaction operates). Let us consider now the stoichiometric coefficients, that are  $S_{\mu,j}<0$ for  substrates and  $S_{\mu,j}>0$ for products. Let us also define the vector $\pmb{g}\equiv\{g_\mu\}$ of the Gibbs energies per mole of species $\mu$. In terms of $\pmb{g}$ we have that $\Delta G_i = S^T_{j,\mu}\;g_\mu$. Now we can discuss the following problem. Given a set of reaction directions $\{ \delta_i\}$ (that have been inferred in a first step of the procedure) determine, \textit{if it exists}, $\pmb{g}$ such that
\begin{equation}
    \Delta_i\equiv -\delta_i\sum_\alpha S_{j,\mu}g\mu\ge 0\,\forall i\;.
\label{eq:con}
\end{equation} 
For fixed $\delta_i$ the solution space is convex. Relaxation methods are a typical and potentially effective choice for solving a problem of this kind. If a solution does not exist the reconstructed network is not consistent and it has, at best, to be cured. 

To solve this problem we have introduced \cite{de2012scalable} an algorithm based on the so called MinOver approach \cite{minover0} which was originally developed for neural network learning. The algorithm starts from a configuration of the $g_\mu$, that is extracted under the probability distribution $P_0(\pmb{g})$. $P_0$ is selected a priori after phenomenological considerations. All the experimental input to the algorithm is indeed in the choice of $P_0$. We assume the simple ansatz 
\begin{equation}
    P_0\left(\pmb{g}\right) = \prod_{\alpha=1}^{M}
    P_0^\mu\left(\pmb{g}\right)\;,
\end{equation}
where $P_0^\mu$ is uniform around the values estimated from experiments with a range also suggested from experimental data.
We now generate a random vector under $P_0\left(\pmb{g}\right)$ and we compute the vector $\pmb{\Delta}=\{Delta_i\}$ from \ref{eq:con}. Let us call $i_0$ the index of the most broken constraint, i.e. let us set
\begin{equation}
    i_0 = \mbox{arg\;} \min_i \Delta_i\;,
\end{equation}
Now if $\Delta_{i_0}\ge 0$ $\pmb{g}$ is a thermodynamically consistent chemical potential, and solves our problem. We can accept it and exit, or look for more solutions (including the one already found) by restarting the algorithm from a different seed. If instead $\Delta_{i_0}< 0$ $\pmb{g}$ we do not have a solution and we update $\pmb{g}$ by setting
\begin{equation}
    \pmb{g}\longrightarrow \pmb{g} - \lambda \delta_{i_0}
    \pmb{S}_{i_0}\,;
\end{equation}
with $\lambda$ an appropriate constant. One iterates till convergence, that is guaranteed (maybe after a very long, unpractical time) if a solution exists.

If the problem has no solutions the assignment of the directions is not consistent. This happens if and only if there is at least one unfeasible loop. The main problem is at this point that is not easy to find a loop that can be, in principle, also very long. We can phrase better the problem by saying that there is an unfeasible loop if there is a set $\mathcal{I}$ of reactions such that a set of positive constants $k_i>0$ exists such that
$$
\sum_{i\in \mathcal{I}} k_i \delta_i S_{\mu,i}=0\;\forall\mu\;.
$$
If there is a loop the algorithm does not converge, since the least satisfied constraint rotates on the loop and does not get fixed. If this is happening we can start from our algorithm to localize and kill not too long loops in the following way. After discarding the first part of the iteration steps (where we typically are in a transient region) we start storing the values of $i_0(t)$ (where $t$ labels the iterations of the procedure), i.e. the value of the most broken constraint at iteration $t$ of the MinOver procedure. Now we look among the reactions appearing in the set of the $\{i_0(t)\}$ for $t>\tilde{T}$, where we can vary the minimum iteration $\tilde{T}$, and we search for loops of length $\mathcal{L}$, starting from   $\mathcal{L}=3$, and increasing it if needed.When we find a loop we change one of the directions and try the MinOver procedure again.
The loops in the dual space can be searched exhaustively or with aid of Monte Carlo methods \cite{DeMartino:2013p4115}

The possibility of improving the reconstruction of a network, making it compatible with thermodynamical basic principles, is important, and our algorithm helps in this direction.

%%%%%%%%%%%%%%%%%%%%%%%%%%%%%%%%%%%%%%%%%%%% ddm
\section*{Flux scales}
It is  interesting to notice  that  scales analysis in metabolic networks  need implementation of thermodynamics constraints  beyond reversibility assignment. This leads to  a geometrically more complex picture of the flux space, but its lack  leads possibly to  wrong conclusions. Previous work on sampling the flux space seemed to show  {\em scale free} distributions \cite{Almaas:2004p4295},  in contradiction with  the  existence of physical limiting factors, e.g. resources availability \cite{monod1949growth} or maximum ribosome elongation rate \cite{bremer1996modulation}.  
As we stressed in the background, models of metabolic networks come with arbitrary bounds on the fluxes, upon which the solution space could depend,  in turn hampering scaling analysis, in particular in presence of the aforementioned unfeasible  loops. Suppose in fact that $f_{i,0}$ is a feasible state and $k_i$ is a solution of (5), the line $f_i(L) = f_{i,0} + k_i s_i L$   verifies the steady state mass balance constraints by construction and it will be inside the polytope   till $L$ possibly reaches the arbitrary  bounds.
\begin{figure}[h!!!!!!!!!!!!!!!!!!!!!!!!!!!!!!!!]
\begin{center}
\includegraphics[width=0.85\textwidth]{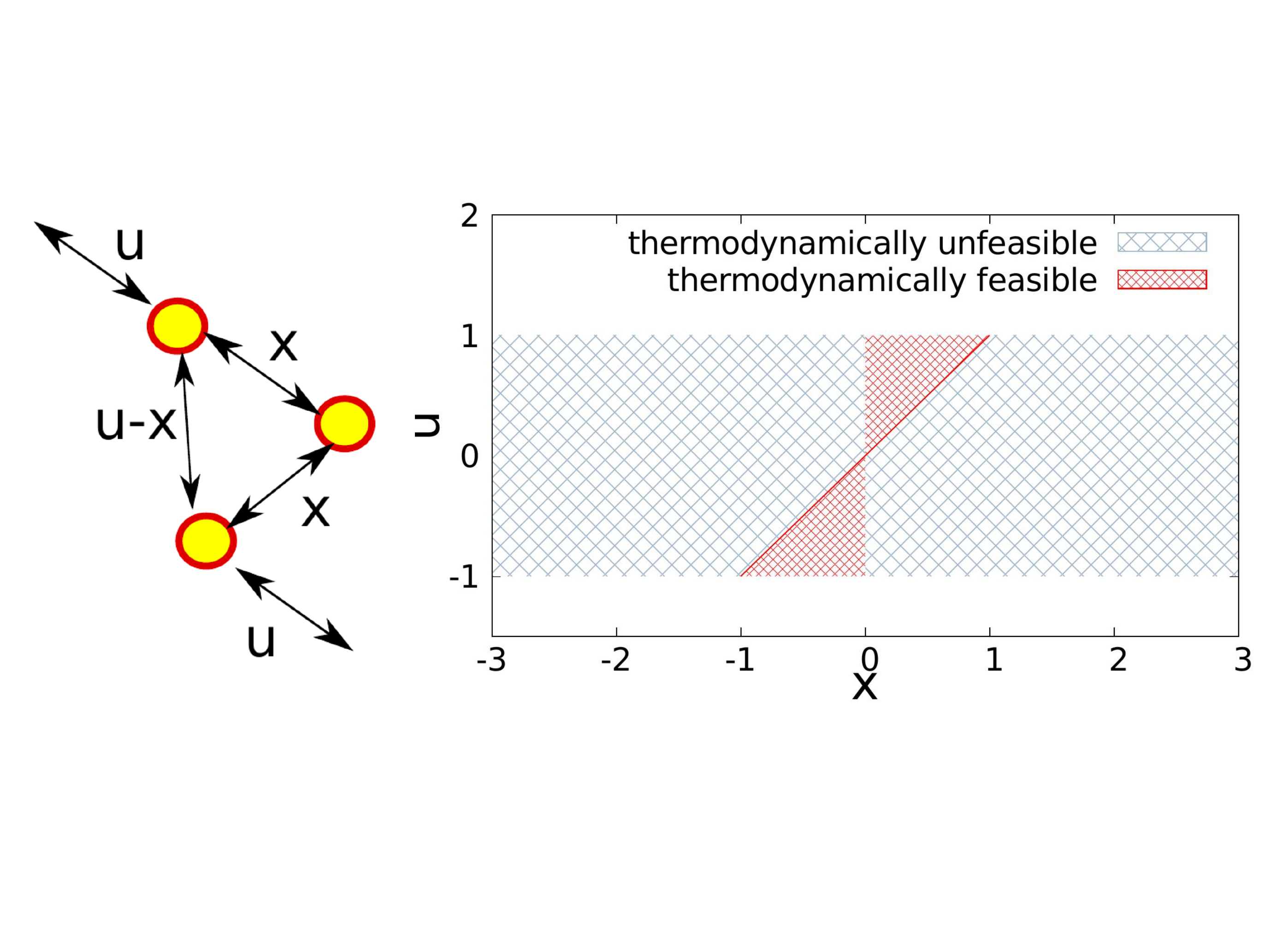}
\end{center}
\caption{A simple three reaction network and its stationary flux space  $(x,u)$, $u\in[-1,1]$ and $x\in[-3,3]$: thermodynamics constraints impose  $ux\geq 0$  $|x|\leq |u|$.}
\end{figure}
A simple illustration is depicted in Fig. 1: $A$ is injected  with rate $u$, it can be transformed either in $B$ and subsequently in $C$ with rate $x$ or directly in $C$ with rate $u-x$, that is consumed with rate  $u$. (if $u<0$,   $C$ is injected and $A$ consumed). 
The variable  $x$ is unbounded unless closed loops are forbidden, leading    to the non linear constraints $u x \geq 0$, $|x|\leq |u|$.
In general thermodynamics forbids the orthants in the flux space that include closed loops:  the remaining feasible space  is  not convex anymore (see Fig. 1), but its scales now  reflect true physical constraints. 
These issues have been studied in a genome scale metabolic network model, specifically the typical steady states of the E Coli metabolic network iJR904 \cite{reed2003expanded}  in a glucose limited minimal medium in aerobic conditions \cite{de2017scales}.  Flux configurations have been uniformly sampled and corrected from unfeasible loops with the  methods described in the previous section.
Results for the  distribution  of flux intensities $|f_i|$  are shown in Fig. 2, before and after correcting for unfeasible loops.
\begin{figure}[h!!!!!!!!!]
\begin{center}
\includegraphics[width=0.85\textwidth]{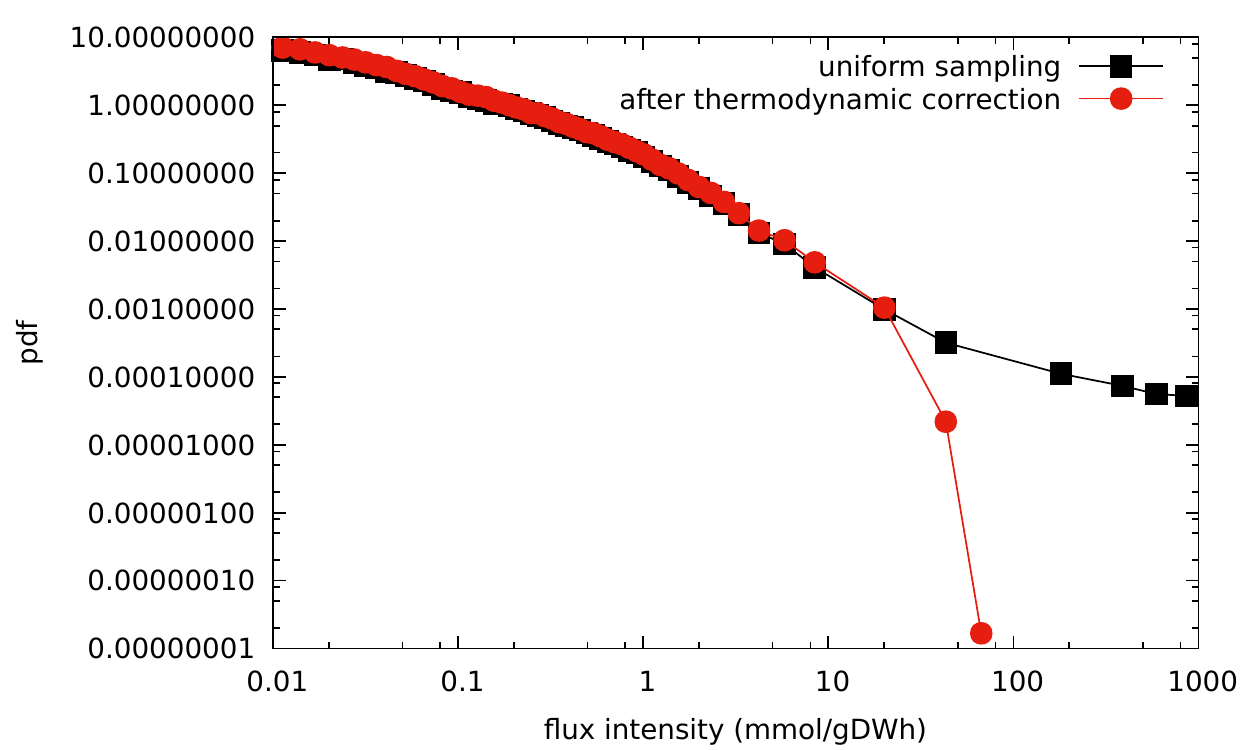}
\end{center}
\caption{ Flux intensity distribution (log-log)  before and after removing thermodynamically unfeasible loops ({\it E.Coli} model iJR904 \cite{reed2003expanded}, $R=10^5$ configurations).}
\end{figure}
The long tail corresponding to the uniform sampling depends on the arbitrary constant fixed for flux bounds. After thermodynamic correction the flux intensity distribution has a cut-off that scales simply with the glucose input \cite{de2017scales}.

%%%%%%%%%%%%%%%%%%%%%%%%%%%%%%%%%%%%%%%%%%%%%% ddm
\section*{Enhanced turnover}
What is the overall turnover time of metabolism?
This question receives useful hints from thermodynamic information even if the knowledge of reactions kinetic is lacking.  In particular   fluctuation analysis upon application of the fluctuation theorem returns a faster picture of metabolism with respect to standard turnover  estimates \cite{de2016genome}.  
In general and formally  relaxation times  are calculated
from a  linear stability analysis of the steady states,
but this requires detailed knowledge of reactions kinetics. 
On the other hand if at least fluxes and concentrations are experimentally known, it is possible to calculate the metabolites turnover times $\tau$, i.e. the ratio between the concentration  $c$ and the net flux of production $P$ (or equivalently consumption $D$, given the steady state), schematically
\begin{equation}
\dot{c} =P -D = 0, \quad \tau =\frac{c}{P} 
\end{equation}
where e.g. the flux $P$   can be calculated from the network and the flux configuration for a  given metabolite $\mu$ ($\theta$ is the Heaviside step function):
\begin{equation}
P_\mu = \sum_{i}\theta(S_{i \mu} v_i)S_{i \mu} v_i
\end{equation}
Such turnover time is the typical time it takes to fully replenish a given metabolic pool.
On the other hand   net fluxes result from the difference between forward and backward contributions
$\nu=\nu^+-\nu^-$,
and the latters can be estimated by the fluctuation theorem if the free energy $\Delta G$ change is known (cit):
\begin{equation}
\frac{v^+}{v^-} = e^{-\Delta G/RT}
\end{equation} 
Upon taking into account the backward contribution the turnover time can be shorter, i.e.  schematically
\begin{eqnarray}
\dot{c} =(P^+ + D^-) -(D^+ + P^-) = 0 \nonumber \\
\tau =\frac{c}{P^+ + D^-} 
\end{eqnarray}
For instance consider  Glucose-6-phosphate in the human red blood cell. This is  produced the  Hexokinase enzyme ($\Delta G_1 \simeq -29$ KJ/mol),  and consumed by  the phosphoglucoisomerase enzyme ($\Delta G_2 \simeq -2.9$ KJ/mol). At $ RT = 2.5$KJ/mol  the  turnover time $\tau_0$ estimated only from net fluxes overestimates the one $\tau$ that takes into account backward contribution  by a factor
\begin{equation}
\frac{\tau_0-\tau}{\tau} \simeq \frac{1}{e^{-\frac{\Delta G_2}{RT}}-1}\simeq 45\%
\end{equation}
Such analysis has been performed  for  the genome scale E.Coli metabolic network iJR904 \cite{reed2003expanded} in a glucose limited minimal medium in aerobic conditions \cite{de2016genome}, returning a faster picture of intermediate metabolism, that we summarize in  table 1  reporting the turnover times estimate from net and total fluxes of the metabolites ruling  the energetics of the network.

\begin{table}[h!!!]
\begin{center}
\begin{tabular}{  | c  | c  |  c | }  
\hline 
Metabolic  & Turnover time  & Turnover time   \\ 
 compound   & estimated from net flux (s) & corrected for fluctuations  (s) \\
\hline 
ATP  &  $2.0\pm0.1$  &  $0.4\pm0.2$  \\
ADP  &   $0.120\pm0.005$ & $0.02\pm0.01$   \\
AMP  &  $0.5\pm0.1$  & $0.11\pm0.06$  \\
NAD  &  $1.1\pm0.1$  & $0.3\pm0.1$  \\
NADH  &  $3.5\pm0.2 \cdot 10^{-2}$  & $1.0\pm0.4 \cdot 10^{-2}$   \\
NADP  &  $1.6\pm0.2 \cdot 10^{-3}$   & $2\pm1 \cdot 10^{-4}$    \\
NADPH  & $9\pm1 \cdot 10^{-2}$    &  $2\pm1 \cdot 10^{-2}$   \\
Glutammate  &  $90 \pm 20$  & $16 \pm 8$   \\
3-Phosphoglycerate  &  $2.0\pm0.2$  & $0.12\pm0.6$   \\
\hline
\end{tabular}
\caption{Turnover times of selected compounds in {\it E Coli} metabolism simulated for a genome scale model in aerobic glucose limited minimal environment, from net fluxes and corrected for fluctuations.}
\end{center}
\end{table}

%%%%%%%%%%%%%%%%%%%%%%%%%%%%%%%%%%%%%%%%%%%%%%% adm
\section*{Variational principles}

An alternative route to implementing thermodynamic feasibility at network scale consists in devising global variational principles ensuring that optimal mass-balanced flux patterns are void of cycles . The simplest such principle is perhaps given by the minimization of the total flux
\begin{equation}
    Q(\mathbf{v})=\frac{1}{N}\sum_{i=1}^N v_i^2~~,
\end{equation}
where it is understood that flux vectors $\mathbf{v}=\{v_i\}$ satisfy the mass balance conditions $\mathbf{Sv=0}$ with pre-defined ranges of variability for fluxes (as well as the additional constraints that may be required on a case by case basis). An argument proving that $Q$ is minimum for flux configurations that are thermodynamically viable (assuming such configurations exist for the network under study) is as follows. Consider a mass-balanced flux configuration $\mathbf{v}$ and assume it contains an infeasible cycle. Such a cycle must be described by a non-zero solution $\mathbf{k}$ to the system
\begin{equation}
    \sum_i \Omega_i^\mu k_i =0~~,
\end{equation}
where $\Omega_i^\mu=-v_i S_i^\mu$. Now consider the flux configuration $\mathbf{w}$ defined by
\begin{equation}
    w_i=v_i+\alpha k_i v_i~~,
\end{equation}
with $\alpha$ a constant. Clearly, if $\mathbf{v}$ is mass-balanced, so is $\mathbf{w}$. However $Q(\mathbf{v})> Q(\mathbf{w})$ provided $\alpha$ is such that $\frac{\partial Q(\mathbf{w})}{\partial\alpha}=0$. In particular, one finds
\begin{equation}
    Q(\mathbf{v})=Q(\mathbf{w})+\frac{\left(\sum_i k_i v_i^2\right)^2}{\sum_i k_i^2 v_i^2}~~.
\end{equation}
In other terms, given a thermodynamically infeasible mass-balanced flux configuration it is always possible to construct another mass-balanced flux configuration whose total flux is lower. In turn, the resulting flux pattern has to be thermodynamically feasible when $Q$ is minimized. 

This idea, originally put forward in \cite{Holzhutter:2004pr}, has been applied in various computational schemes for genome-scale metabolism, such as pFBA (parsimonious Flux Balance Analysis) \cite{lewis2010omic}, CycleFreeFlux \cite{desouki2015cyclefreeflux}, the global method to remove infeasible cycles from NESS flux configurations introduced in \cite{DeMartino:2013p4115}. Clearly, it is useful in practice whenever one is interested in finding a single feasible flux pattern as long as the minimum of $Q$ lies within the solution space defined by mass balance constraints. A more integrative principle has been proposed in \cite{fleming2012variational}. It is most easily expressed by distinguishing forward ($F$) and reverse ($R$) directions for each flux, so that the net flux $v_i$ can be written as $v_i=v_{i,F}-v_{i,R}$ (with $v_{i,F}\geq 0$ and $v_{i,R}\geq 0$), as well as exchange fluxes corresponding to sources or sinks of the reaction network ($v_{i,E}$). In brief, it states that the triplet $(\mathbf{v}_F^\star,\mathbf{v}_R^\star,\mathbf{v}_E^\star)$ satisfies a Flux Balance Analysis-like problem (i.e. is mass balanced and maximizes a given linear objective function) thermodynamically, provided it minimizes the functional
\begin{equation}
    \mathcal{F}(\mathbf{v}_F,\mathbf{v}_R)=\mathbf{v}_F\,\cdot\,\left[\log(\mathbf{v}_F+\mathbf{c}-\mathbf{1})\right]+\mathbf{v}_R\,\cdot\,\left[\log(\mathbf{v}_R+\mathbf{c}-\mathbf{1})\right]
\end{equation}
subject to 
\begin{equation}\label{constraint}
    \mathbf{S}_I(\mathbf{v}_F-\mathbf{v}_R)+\mathbf{S}_E\mathbf{v}_E^\star =\mathbf{0}~~,
\end{equation}
where $\mathbf{S}_I$ and $\mathbf{S}_E$ stand for the intracellular and exchange parts of the stoichiometric matrix, $\mathbf{c}$ is a generic vector in $\mathbb{R}^N$ and $\mathbf{1}$ is the vector with all entries equal to 1. In other terms, if $\mathbf{v}_E^\star$ is the vector of optimal exchange fluxes for the solution of an FBA problem, one can obtain a thermodynamically viable solution to the same FBA problem by minimizing $\mathcal{F}$. Importantly, the vector $\mathbf{g}$ of chemical potentials are related to the vector $\boldsymbol{\lambda}$ of Lagrange multipliers enforcing (\ref{constraint}) by
\begin{equation}
    \mathbf{g}=-2RT\boldsymbol{\lambda}~~.
\end{equation}
This result follows from standard convex analysis (see \cite{fleming2012variational} for details) and benefits from the standard conceptual and computational advantages of convex optimization problems with linear constraints (uniqueness of solution, efficient computational implementation). In addition, it is fully generic, in the sense that each thermodynamically viable mass-balanced flux vectors must minimize $\mathbf{F}$ for a certain choice of $\mathbf{c}$. It therefore provides a rather transparent description of thermodynamic feasibility (as far as optimality is concerned). On the other hand, the existence of a free parameter constitutes a limitation, at least in part. The most serious drawback, in our view, however derives from the fact that the above formulation does not allow to account for explicit constraints on {\it net} fluxes (as discussed in \cite{fleming2012variational})

Thermodynamic arguments have also inspired different types of variational principles that effectively extend the reach of flux-based models by allowing to account for concentrations. For instance, in Ref. \cite{de2014inferring} feasible NESS are assumed to minimize the function
\begin{equation}\label{Hu}
    H=\sum_{\mu\,\in\,{\rm Ext}}\frac{(u^\mu)^2}{c^\mu_{\rm ext}}
\end{equation}
over exchange fluxes $\mathbf{u}$ and intracellular fluxes $\mathbf{v}$, subject to
\begin{gather}
    \mathbf{Sv=u}
\end{gather}
and with prescribed bounds of the form $u^\mu\in[u^\mu_{\min},u^\mu_{\max}]$ and $v_i\in[v_{i,\min},v_{i,\max}]$ for each extracellular compound $\mu$ and each intracellular reaction $i$. The explicit dependence of  $H$ on external concentrations (considered as fixed parameters) makes it possible to use the above principle to infer intracellular reaction rates given the levels of a set of extracellular metabolites. On the other hand, the minimization of (\ref{Hu}) does not ensure that the resulting flux pattern is thermodynamically viable, unless in specific cases. The reader is referred to \cite{DeMartinoPLoSONE_2012} for details.

%%%%%%%%%%%%%%%%%%%%%%%%%%%%%%%%%%%%%%%%%%%%%%% adm
\section*{Conclusions}

Integrating thermodynamics with genome-scale biochemical  reconstructions is perhaps the central theoretical open challenge of metabolic network modeling. Basically, two classes of approaches are currently being attempted that can mutually benefit from each other. On one hand, novel empirical data and biochemical methods are employed to estimate more accurate standard free energies for reactions and chemical potentials for metabolites \cite{Jankowski:2008bh,noor2013consistent }. Having better estimates of such quantities is crucial, especially if they cover a larger part of the reactome and of the metabolome than the ones currently available. On the other hand, optimization principles based on different physico-chemical arguments can be used to obtain approximate (but genome-scale) estimates, whose accuracy depends on the prior biochemical information as well as on the underlying assumptions. The key issue to be faced here is computational, relating both to the scalability of the algorithms required for the study of genome-scale networks and to the fact that thermodynamics may require the study of non-convex optimization problems. At the same time, ongoing  work is uncovering how thermodynamics constrains flux patterns at steady state starting from the kinetics of individual enzymes, with results which suggest that the problem of computing thermodynamic potentials described here might have to be modified as more is known about individual reaction mechanisms, at least to some degree \cite{noor2014pathway}. In this respect, it is the convergence of novel statistical physics \cite{polettini2014irreversible, seifert,wachtel2018thermodynamically, rao2018conservation }, biochemical and algorithmic ideas that will likely provide the tools to effectively tackle this challenge.

\bibliographystyle{unsrt}
\bibliography{reference}
%\blankpage
%\printindex[aindx]                 % to print author index
%\printindex                         % to print subject index

\end{document}